\documentclass[twocolumn,prb,superscriptaddress,aps,10pt]{revtex4-1} 

\usepackage{amsmath}  
\usepackage{amsfonts} 
\usepackage{graphicx} 
\usepackage{hyperref}
\usepackage{enumitem}
\usepackage{minibox}
\usepackage{color}

\newcommand{\pcubed}{${\rm P}^3$~}

\begin{document}


\title{\pcubed: A Practice Focused Learning Environment}

\author{Paul W. Irving}
\email{Corresponding author: pwirving@msu.edu} 
\affiliation{Department of Physics and Astronomy, Michigan State University, East Lansing, MI 48824, USA}
\affiliation{CREATE For STEM Institute, Michigan State University, East Lansing, MI 48824, USA}

\author{Michael J. Obsniuk}
\affiliation{Department of Physics and Astronomy, Michigan State University, East Lansing, MI 48824, USA}

\author{Marcos D. Caballero}
\affiliation{Department of Physics and Astronomy, Michigan State University, East Lansing, MI 48824, USA} 
\affiliation{CREATE For STEM Institute, Michigan State University, East Lansing, MI 48824, USA}



\date{\today}

\begin{abstract}

There has been an increased focus on the integration of practices into physics curricula, with a particular emphasis on integrating computation into the undergraduate curriculum of scientists and engineers. In this paper, we present a university-level, introductory physics course for science and engineering majors at Michigan State University (MSU) called \pcubed (Projects and Practices in Physics) that is centered around providing introductory physics students with the opportunity to appropriate various science and engineering practices. The \pcubed design integrates computation with analytical problem solving and is built upon a curriculum foundation of problem-based learning, the principles of constructive alignment and the theoretical framework of community of practice. The design includes an innovative approach to computational physics instruction, instructional scaffolds, and a unique approach to assessment that enables instructors to guide students in the development of the practices of a physicist. We present the very positive student related outcomes of the design gathered via attitudinal and conceptual inventories and research interviews of students' reflecting on their experiences in the \pcubed classroom. 

\end{abstract}

\maketitle 

\section{Introduction}\label{sec:intro}

The appropriation of scientific practices has become a focus of learning goals both in the K-12 and undergraduate arena. In 2012, the National Research Council released ``A Framework for K-12 Science Education: Practices, Crosscutting Concepts, and Core Ideas" (The Framework), \cite{NRC2011} which highlighted science practice as an important focus for STEM education. Although an introductory college science course is a significantly different context than the target audience of the K-12 community, it has been argued that the framework can and should be applied to introductory science courses. \cite{Cooper2015} The Framework discusses eight science and engineering practices that are critical for future science curricula. \cite{NRC2011} Two critical practices -- ``developing and using models" and ``using mathematics and computational thinking" -- are essential to physics instruction as they relate to computational modeling -- the ``third leg'' of modern science and engineering.\cite{denning2007,kirk2016} 

Modern scientists increasingly rely on computational modeling (the use of a computer to numerically solve, simulate, visualize, and explain physical phenomenon) to explore and to understand the natural world; however, computational modeling is largely ignored in most introductory college physics courses.
As computers have become more powerful, 21$^\mathbf{ st}$ century scientists can model complex physical systems in ways that were impossible a decade ago. 
For example, scientists model complex systems to gain insight into that system's properties (e.g., understanding recurrence in turbulent flows \cite{cvitanovic2013}), simulate impossible experiments numerically (e.g., predicting the dynamics of galactic mergers \cite{kim2009}); and reduce mountainous experimental data to a sensible size (e.g., extracting significant events from elementary particle collisions \cite{duhrssen2004}).
Yet, as computer {\it usage} has grown amongst both faculty and students, most introductory STEM courses have failed to introduce students to computation's problem-solving powers. 
While some attempts have been made to incorporate computational modeling into introductory STEM courses (e.g., in physics \cite{MacDonald1988,Schecker1994,Roos2006,Chabay2008}), evidence-based instructional design and practice have not yet been incorporated. 
So, while we know computational modeling is the tool of 21$^\mathrm{ st}$ century science and engineering, we have little evidence for how to teach it well.

In this paper, we discuss a university-level, introductory physics course for science and engineering majors at Michigan State University (MSU) called \pcubed (Projects and Practices in Physics) that is centered around providing introductory physics students with the opportunity to appropriate various science and engineering practices, with a particular focus on computational modeling. In Sec.~\ref{sec:motiv}, we provide the design principles of \pcubed including how we leveraged constructive alignment\cite{Biggs2011} in our design.

\section{Motivation \& Philosophy}\label{sec:motiv}

\pcubed was designed using the principle of constructive alignment,\cite{Biggs2011} which argues that designers focus on the learning goals of the learning environment first, and then reverse engineer the assessment and instruction to align with the identified learning goals. 
Based on the arguments presented in The Framework,\cite{NRC2011} we decided that the \pcubed learning environment should emphasize science practice with a particular focus on computational modeling. 
Drawing from the recommendations from the literature,\cite{CommitteeontheStatus:2012wj,NRCper:2013} we selected an instructional model that emphasized self-directed group learning, which positioned the instructional staff as facilitators of learning. 
To ensure the alignment of the instructional approach, instruction on computational modeling also had to emphasize self-directed learning.
The \pcubed appraoch to computational instruction makes use of a non-traditional perspective, which will be outlined in the section on ``minimally working programs''.\cite{Weatherford2011,Lunk2012,Caballero:2012hu,Caballero:2012vb}
It is important to emphasize that it is not our intention that students become fluent in a computational language (in the case of \pcubed: python).
Instead, \pcubed is an opportunity to introduce students to the utility of computational modeling and some basic programming structures such as iterative loops.\cite{Chabay2008} 

However, we recognize computational modeling is an important practice for students to engage in preparation for their future career. 
Part of our motivation for providing students with the opportunity to participate in computational modeling is that it is a central practice for both the science and engineering communities that our student population in \pcubed are on a trajectory to join. 
The term ``central practice" is derived from the communities of practice framework \cite{Wenger1998} that argues that a student will develop an identity in a discipline by engaging in central practices of that discipline while being guided by central members of that discipline.\cite{Barab2002,Irving2014,Irving2016} 
\pcubed is an introductory calculus-based mechanics course at MSU for students whose majors tend to be focused in science and engineering. 
Research has demonstrated that engaging in more central practices of a physics community like undergraduate research can help students identify themselves as physicists.\cite{Hunter2007,Irving2015}
In turn, developing an identity in a particular discipline can help students to persist with said chosen discipline.\cite{Pierrakos2009} 
However, undergraduate research and, in turn, the opportunity to engage in the authentic central practices of science and engineering communities are challenging to provide equally to all students at the introductory level. 
Despite this difficulty, the \pcubed design aims to engage students in more authentic practices of science and engineering by encouraging students to feel a bit more central in their science and engineering communities. 
Therefore, our design is focused on the appropriation of what we refer to as ``pathway practices," which are base practices ubiquitous to both science and engineering communities that provide a foundation for future central practices. 
For example, we have designed \pcubed  so that it is focused on developing students' ability to work successfully in a group. This pathway practice forms the basis for central practices in science and engineering such as collaborating on multi-faceted research experiments and/or working as part of a design team. 
For a full list of the practices incorporated into our design, please refer to Sec.~\ref{sec:four}. 

The instructional model that we present in this paper has many moving parts, which were decided on due to both contextual constraints of our institution and our chosen emphasis on particular learning goals. 
Full adoption of this design might be challenging in different contexts, but by outlining the constituent parts and the decisions behind their design and inclusion, we aim to afford the opportunity for the adoption of parts of this design. 
We are presenting a model for courses that wish to emphasize science practice.





\section{Projects \& Practices in Physics}\label{sec:act}

Like many universities, the traditional, introductory mechanics curriculum at MSU follows a historical canon of topics, which limits students' experiences to 17$^\mathrm{th}$ and 18$^\mathrm{th}$ century physics. 
The organization of the standard curriculum fails to emphasize the interconnectedness and ubiquitous nature of fundamental physical principles (e.g., conservation of momentum and energy), and, instead, emphasizes the use of laundry lists of special case formulas -- contrary to the integrated nature of science and the arguments presented by The Framework.\cite{NRC2011}
\pcubed is a transformed, introductory, calculus-based mechanics course for science and engineering majors that is grounded in evidence-based pedagogy (including peer discussion and conceptual \& reasoning-focused homework) and is built on Matter \& Interactions (M\&I), a fully-developed, extensively-tested, introductory physics curriculum.\cite{Chabay2002,Kohlmyer:2009ib,Caballero:2012ee}
While M\&I was designed and developed before The Framework, the M\&I curriculum and thus, \pcubed, emphasizes a number of the critical aspects of The Framework.
For example, \pcubed is organized around core fundamental principles (conservation of momentum, energy, and angular momentum) meant to emphasize a first-principles approach to problem-solving,\cite{Chabay2004} which are akin to Core Ideas in The Framework.
Additionally, \pcubed places an emphasis on modeling and model building by presenting microscopic models to help students explain macroscopic phenomenon,\cite{Chabay1999} and introducing modern problem-solving tools (i.e., computational modeling),\cite{Chabay2008} which represent a commitment to Cross-Cutting Concepts and Science and Engineering Practices respectively.
The self-directed group nature of the \pcubed learning environment provides an opportunity to emphasize a number of these science practices.
For our design, we have chosen to anchor the course and to build opportunities for learning around ``modeling,'' which we describe below. 

\section{The \pcubed  Approach to Modeling}\label{sec:model}

The \pcubed course engages students in modeling real-world situations using a group-oriented pedagogy -- emphasizing critical aspects of professional science and engineering practice. 

Because the separation of lecture sections and lab courses is a larger contextual issue at MSU, students cannot work with laboratory equipment or conduct experiments in a lecture section. 
Hence, students experience with modeling in the \pcubed course makes use of complex, real-world problems for which they develop analytical and computational solutions.
Typically students will solve one of these complex, real-world problems in each two-hour class meeting (twice per week). 
Of the roughly 30 such problems, 7 are computational modeling problems (described in Sec.~\ref{sec:example}).
Students negotiate their approach and solutions in groups of 4, which was a deliberate decision in order to provide students with the opportunity to learn about group dynamics.
These discussions are minimally facilitated by course staff (faculty, graduate teaching assistants, and undergraduate learning assistants who are collectively called ``tutors''); we push students to answer each others' questions and build their complete understanding of the solution together.
Once students develop a solution to the problem, they present it to their tutor.
We probe students more deeply about their solutions using tutoring questions that facilitate student understanding of the conceptual underpinnings of the problem.

\subsection{Modeling Framework} 

\begin{figure}[ht]
\centering
\includegraphics[width=1\linewidth]{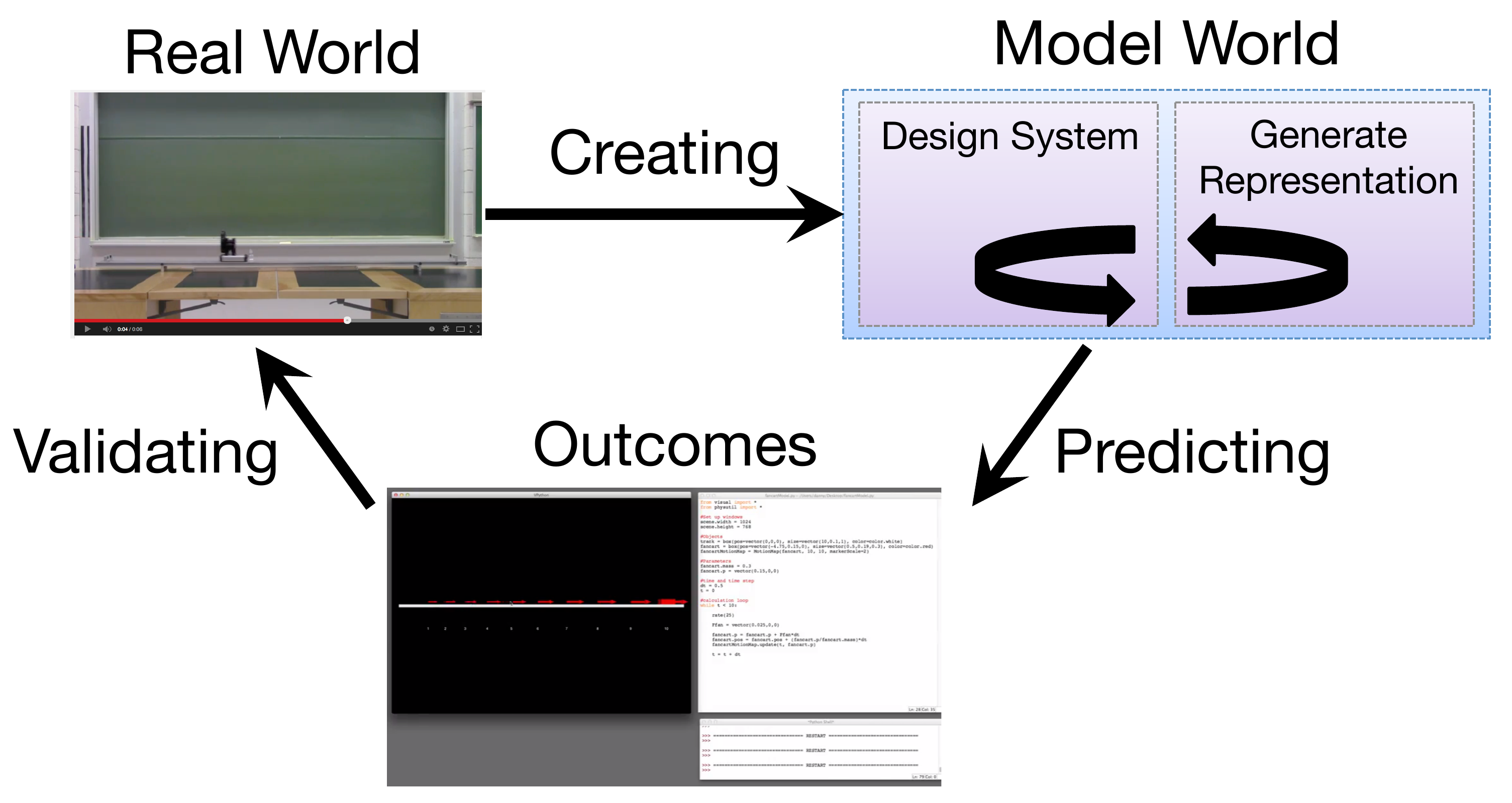}
\caption{Modeling Framework for Projects and Practices in Physics around which student engagement is organized.}\label{fig:modeling}
\end{figure}

Students are guided in their modeling work through their use of the Modeling Framework (Fig.~\ref{fig:modeling}), which was developed by consulting the literature on modeling \cite{Hestenes1992,Hestenes1987,Wells1995,Hoskinson2014}, discussing the development team's experience with fundamental physics research, and considering the larger contextual limitations of MSU's introductory physics program (i.e., the divorce of lecture and lab).
Students begin their work by considering a Real World situation -- this is the complex, real world problem that is posed to their group. From this problem statement, students must create a Model World \cite{Hoskinson2014} in which they will develop their model of the system.
Here, students must provide an explicit articulation of the assumptions and approximations they are making while they attempt to represent different aspects of the problem (Design System \& Generate Representation).

Arguably, much of the discussion and negotiation process within the group occurs here while students (iteratively) design and represent their system in increasingly finer scales until they are confident that they can generate a prediction (or explanation).
Throughout this process, tutors facilitate student work by addressing group concerns, by helping groups construct their understanding through discussion among group members, and, if needed, by asking leading questions and tutoring groups on concepts.
Once their model is clearly articulated, students use their representations (graphs, equations, diagrams) to predict any Outcomes. 
Finally, those Outcomes are validated against Real World results, which may be data, other predictions, or expectations from their own experience in the world.
Through this process, students are developing a {\it solution} to the problem, not simply finding an {\it answer}.
Their solution consists of the work through their entire Modeling process, which is discussed with course staff once students feel they have sufficiently completed a problem.
By focusing on a solution and the modeling process, our design intends to make the activities authentic. 

\section{\pcubed Learning Goals}\label{sec:goals}

While modeling is the framework that supports students' activities, in keeping with the principle of constructive alignment,\cite{Biggs2011} the particulars of the course were developed from the learning goals.
\pcubed uses a number of learning goals to direct pedagogy.  
These goals have been generated from two broad categories of focus: the practices we expect students to engage in and the content we hold them responsible for.  
Within each of these two broad categories the focus leans towards two sub-categories: the fundamental physics principles and the computational implementation of those principles.

\subsection{Practices}

A selection of the scientific practices inspired by The Framework \cite{NRC2011} for which we have developed learning goals are enumerated below in no particular order:  

\begin{enumerate}[noitemsep,nolistsep]
\item[P1.] developing and using models,
\item[P2.] planning and carrying out investigations,
\item[P3.] analyzing and interpreting data,
\item[P4.] constructing explanations,
\item[P5.] and engaging in argument from evidence.
\end{enumerate}  

\noindent These scientific practices were used to inform not only the shared learning goals but also the materials (i.e., projects, exams, homework) we use to achieve those goals.

Developing and using models (P1) is one of the scientific practices used heavily on both analytic and computational problems.  
Whether their models be mathematical or computational, we expect students to work together in groups to develop the model and to utilize that model in further investigations.  
As an example, P1 was used to develop learning goals such as {\it analyze and evaluate data to explain the motion of objects and the responsible interactions} as well as {\it evaluate the applicability/limitations of models and the validity of predictions for different types of motion.}
This type of scientific practice (P1) and the associated learning goals were further used to generate the type of in-class project explicated in Sec.~\ref{sec:example}.

\subsection{Content}\label{subsec:content}

Within the broad content goals of \pcubed are many ideas that we expect students to grapple with both individually and within their group.  
A selection of the previously mentioned sub-categories that we focus on are enumerated below.  
The first set focus on the fundamental physics principles, which are akin to Core Ideas from The Framework: 
\begin{enumerate}[noitemsep,nolistsep]
\item[F1.] macroscopic phenomena are the result of atomic interactions,
\item[F2.] forces external to a system can change the system's momentum,
\item[F3.] work done on or by a system and heat exchanged with the system's surroundings can change the system's energy, and
\item[F4.] torques external to a system can change the system's angular momentum.
\end{enumerate} 
The second set focus on the computational implementation of those principles, which takes full advantage of the power of computation: 
\begin{enumerate}[noitemsep,nolistsep]
\item[C1.] computation can be used to predict the otherwise intractable dynamics of real-world phenomena, and
\item[C2.] computation can be used to generate dynamic and graphic representations of real-world phenomena.
\end{enumerate}  
Each of these sub-categories of focus were used to inform the shared learning goals and materials used in \pcubed.


For example, one learning goal that is presented to students and emphasizes F2 is: {\it apply the momentum principle ($\Delta\vec{p}=\vec{F}_{\rm net}\Delta t;\,d\vec{p}=\vec{F}_{\rm net}dt$) analytically to predict the motion or determine the properties of motion/net force acting on a single-particle system where the net force is a constant vector (e.g., due to the near Earth gravitational force).}
This goal focuses on the application of a fundamental physics principle (\textit{the Momentum Principle}, $\vec{p}_f = \vec{p}_i + \vec{F}_{\rm net}dt$) in order for students to better understand the relationship between dynamics and motion.  
This type of reasoning is used extensively in parts A and B of the type of in-class project described in Sec.~\ref{sec:example}.


The computational implementation of the learning goal presented above focuses on C1: {\it apply the momentum principle ($\Delta\vec{p}=\vec{F}_{\rm net}\Delta t;\,d\vec{p}=\vec{F}_{\rm net}dt$) iteratively/computationally to predict the motion or determine the properties of motion/net force acting on a single-particle system where the net force is not constant (e.g., due to spring-like restoring forces or dissipative drag forces).} 
This goal also focuses on the application of the Momentum Principle, yet takes full advantage of the computational power afforded to the students.  
The generalizable nature of computational methods shows up in part C of the in-class project described in Sec.~\ref{sec:example} where students include a velocity-dependent drag force.

%
%

\section{Example Project}\label{sec:example}

In order to best illustrate the scientific practices students engage with in \pcubed, we present a typical problem that requires both analytical and computational techniques over the course of a week (two in-class meetings of two hours each).  
During the second week of class, students are learning how to predict the motion of point particle systems using Newton's Second Law (\textit{the Momentum Principle}, $\vec{p}_f = \vec{p}_i + \vec{F}_{\rm net}dt$) in a project called Escape from Ice Station McMurdo.

\begin{figure}[t]
\centering
\includegraphics[width=1\linewidth]{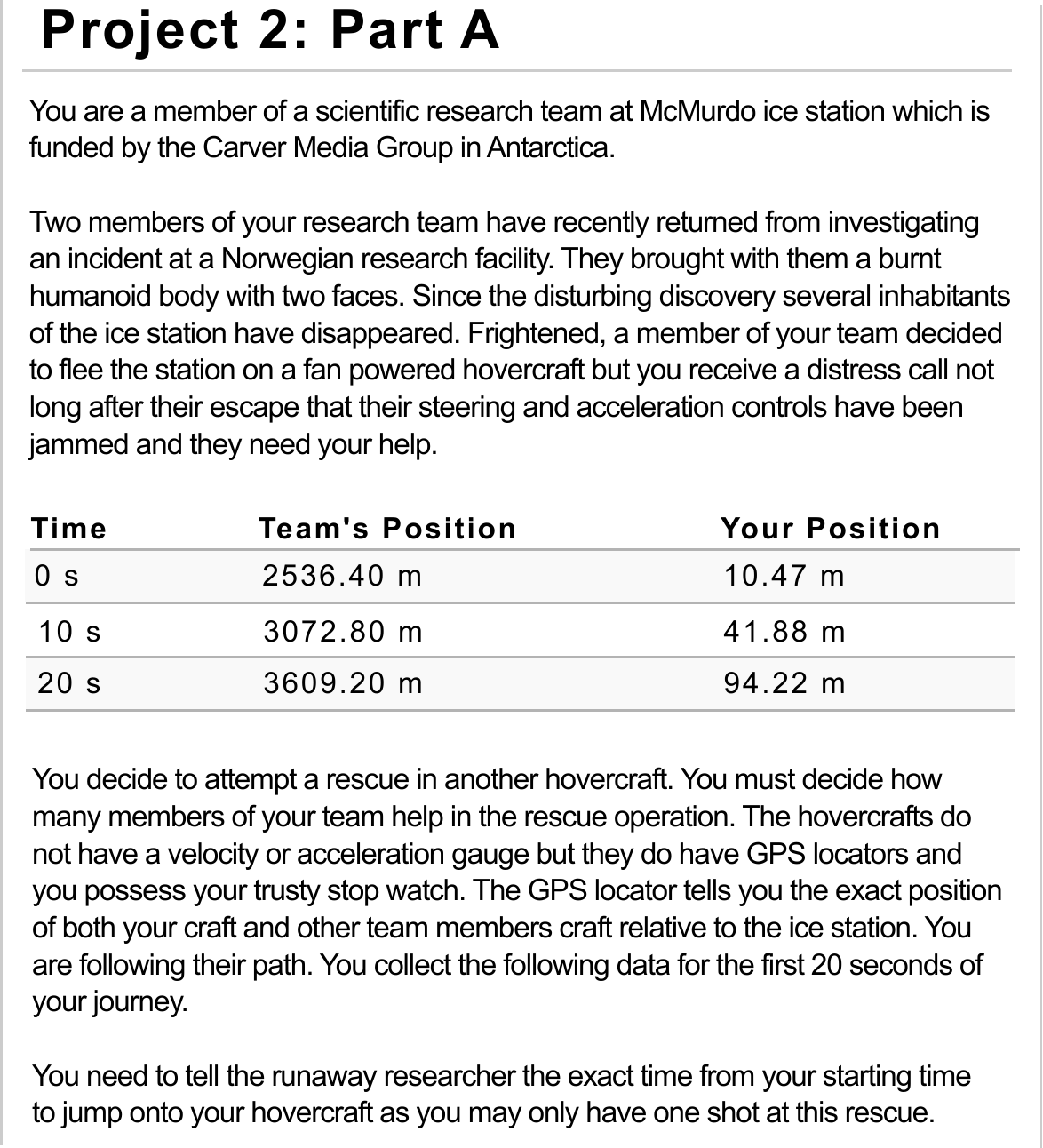}
\caption{The first part of a project where students are asked to model the motion of an object given position vs time data.}\label{fig:p2pa}
\end{figure}

\begin{figure}[t]
\centering
\includegraphics[width=1\linewidth]{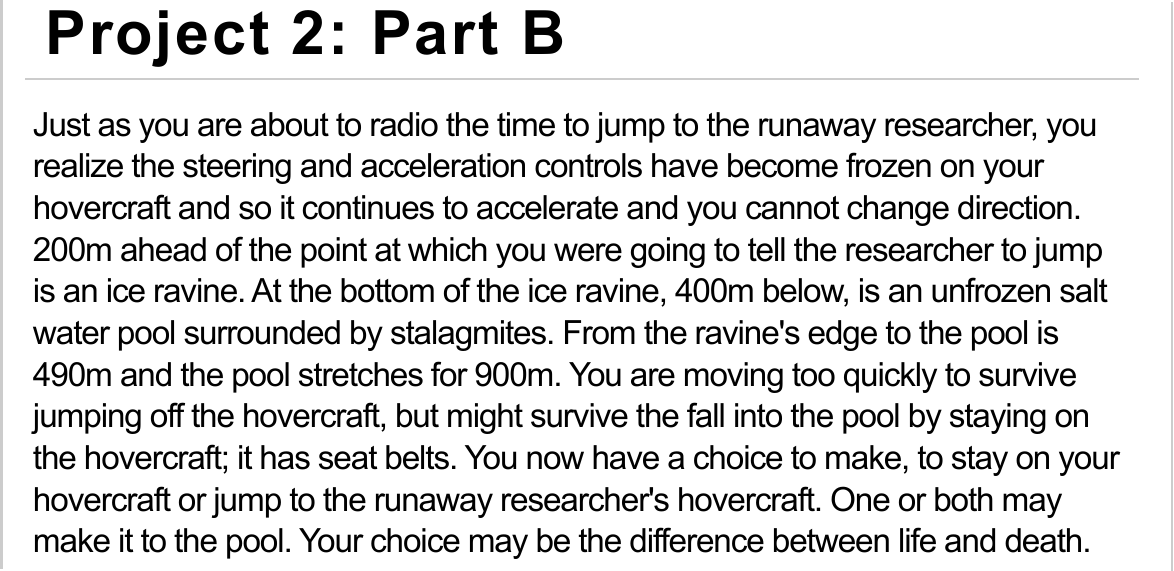}
\caption{The second part of a project where students are asked to model the motion of an object under free-fall conditions.}\label{fig:p2pb}
\end{figure}

\begin{figure}[t]
\centering
\includegraphics[width=1\linewidth]{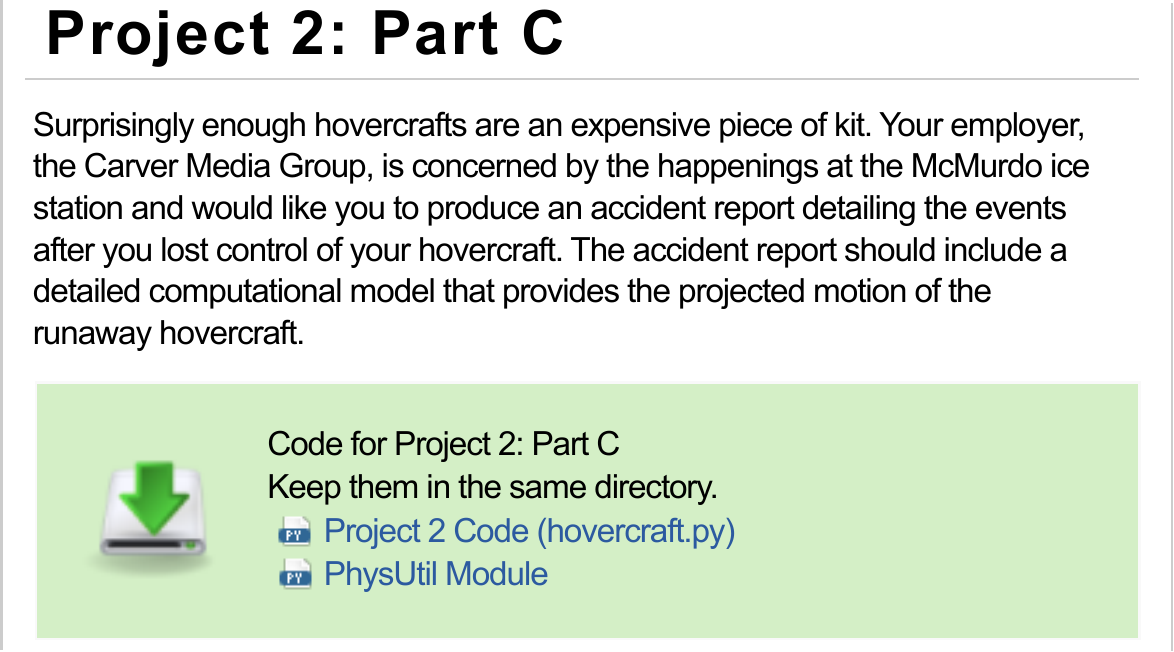}
\caption{The third part of a project where students are asked to model the motion of an object computationally.}\label{fig:p2pc}
\end{figure}

In the first part of this three-part problem (Fig.~\ref{fig:p2pa}), students work with position vs time data to model analytically the motion of two hovercrafts as they race across an ice field -- determining the net force from this data and using appropriate models (constant velocity vs constant force) to predict when the hovercrafts will be at the same location.  
In the second part of Escape from McMurdo (Fig.~\ref{fig:p2pb}), students find that the controls of both hovercrafts are frozen and they are heading towards a cliff.  
They must determine which hovercraft to board in order to survive the fall (there's a salty unfrozen pool at a specific distance from the bottom of the cliff).  
The generality and ubiquity of the Momentum Principle is highlighted in the third part of Escape from McMurdo (Fig.~\ref{fig:p2pc}) where the students computationally model the motion of the hovercrafts including air drag. 

As written above, the Momentum Principle can be used iteratively (through Euler-Cromer integration) to predict the motion of most systems accurately.\cite{Cromer:1981vm}  
It turns out both hovercrafts will land safely when modeled without air drag, but students are told that one hovercraft crashes (it lands short). 
They must file an accident report, which includes a simulation of the accident, to explain how the accident occurred. 

%
%

\section{Learning Scaffolds}\label{sec:scaf}

These weekly projects (Figs. \ref{fig:p2pa}--\ref{fig:p2pc}) are multifaceted and complex.
Hence, a number of different learning scaffolds have had to be developed and put into place to support student learning.  
These scaffolds are meant to support the students in not only the physics concepts they must use, but also in the scientific practices that they must engage in.  
Below, we expound on four such learning scaffolds, which support students' modeling practices (Sec.~\ref{sec:modscaffold}) and their acquisition of course content (Sec.~\ref{sec:contentscaffold}).

\subsection{Modeling Scaffolds}\label{sec:modscaffold}

\subsubsection{The Four Quadrants}

\begin{figure}[t]
\centering
\includegraphics[width=1\linewidth]{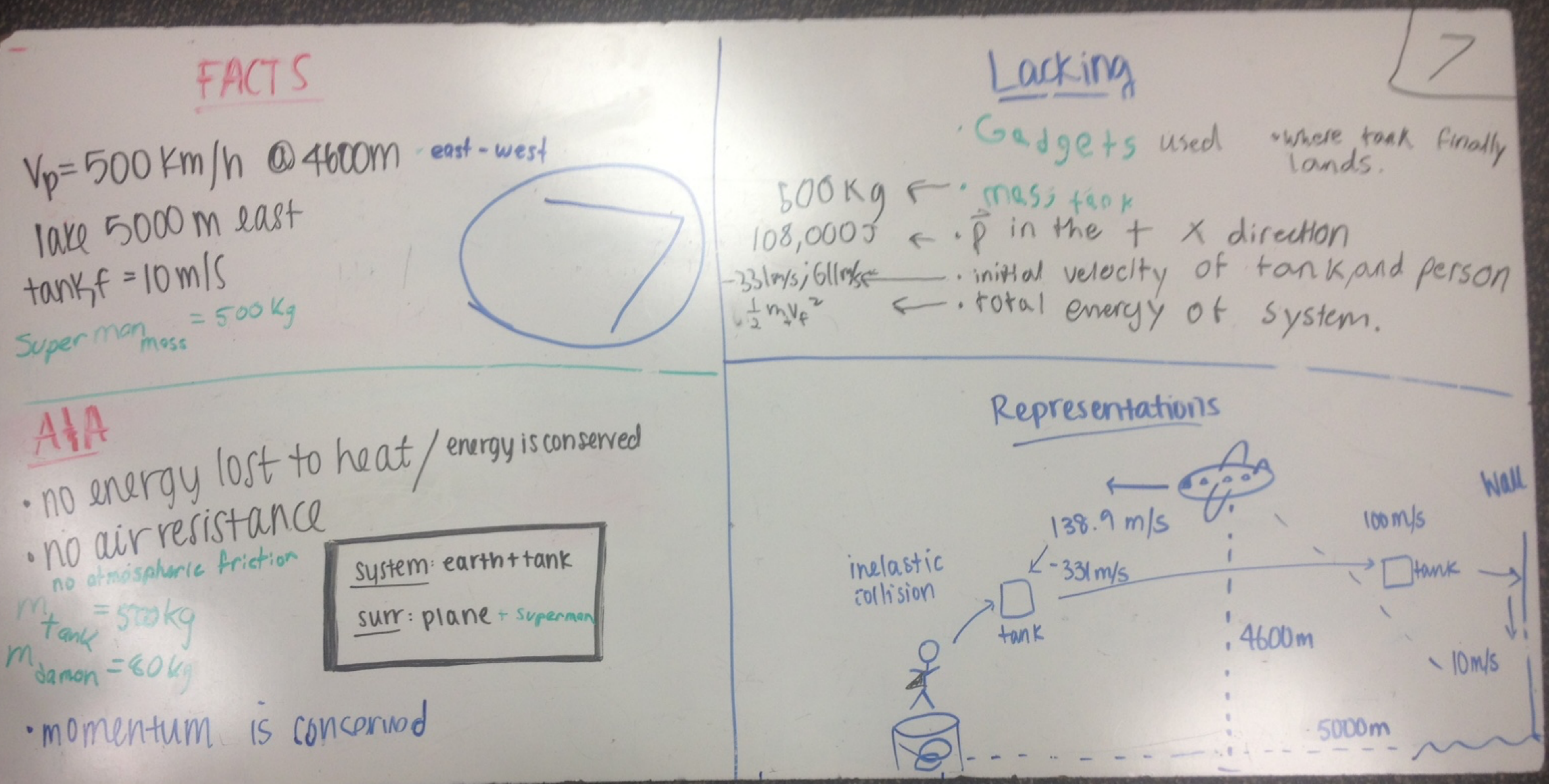}
\caption{An example of a white board with a completed four quadrants}\label{fig:quadrants}
\end{figure}

To scaffold the modeling process (Fig.~\ref{fig:modeling}), we have introduced a conceptual tool -- the Four Quadrants.  
The Four Quadrants provide a designated location for students to record, to reference, and to update collectively agreed upon information during the modeling process.  
In a sense, the Four Quadrants form the basis of the model, and we provide a designated white board to each group for just this purpose.  
Using the Four Quadrants (Fig.~\ref{fig:quadrants}), 
\begin{itemize}[noitemsep,nolistsep]
\item students identify the {\bf Facts}, which are presented in the problem statement;
\item they determine what is {\bf Lacking} from the information they have or can obtain easily;
\item they discuss and negotiate the {\bf Assumptions \& Approximations} they are making; and
\item they provide {\bf Representations} of the problem, which may include diagrams, graphs, or equations.
\end{itemize}

The Four Quadrants are displayed publicly so that each member of a group can easily see and modify the agreed upon information.
Furthermore, this public display allows a group's tutor to ``ping'' the group to see where they are in the solution process.  That is, without needing to disrupt the solution process, a tutor can get a well-rounded idea of what the group is thinking and where the group is heading.  
This is particularly useful as tutors are working with multiple groups at any given time.

\subsubsection{Minimal Working Programs}

A number of problems in \pcubed require that motion be modeled computationally and that students generate dynamic plots (e.g., Fig.~\ref{fig:p2pc}).  
To keep up with the increasing complexity of the problems, students must learn to write short programs using VPython.  
However, less than 10\% of students taking \pcubed have any significant prior computational experience.  
We have extended the work of Weatherford\cite{Weatherford2011} and Lunk,\cite{Lunk2012} who introduced the concept of ` `Minimal Working Programs'' (MWPs) to scaffold student sense-making about computing, and to develop a unique, group-oriented, inquiry-based instructional model for computation.  
This use of MWPs not only encourages sense-making, but also helps to assuage the anxiety many students have about engaging in computation for the first time.  
In this way, we further scaffold the (computational) modeling process.

\begin{figure}[t]
\centering
\includegraphics[width=1\linewidth]{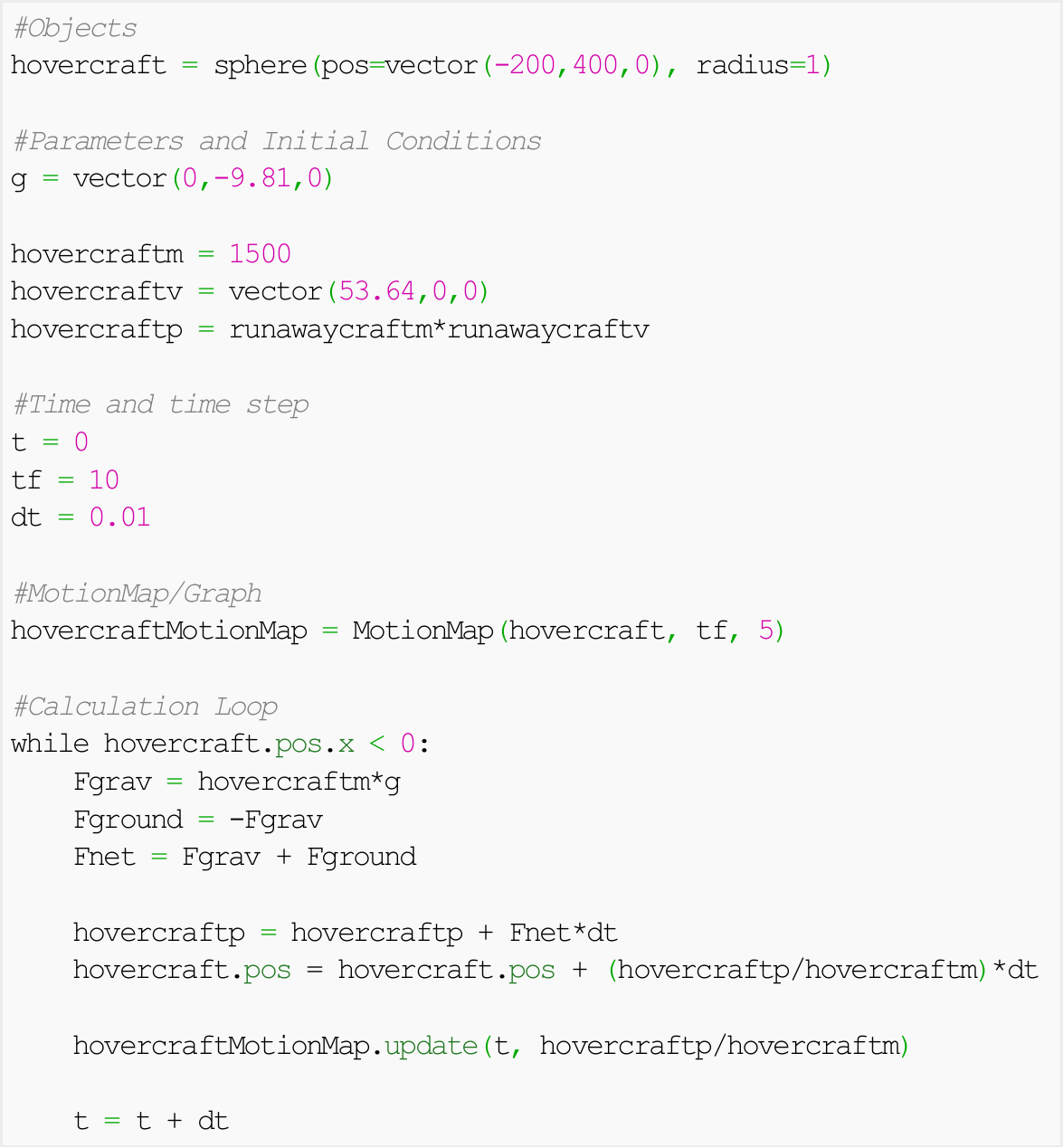}
\caption{Minimally working program that students are provided with for the hovercraft project.}\label{fig:code}
\end{figure}

In \pcubed, students are given no explicit instruction on VPython.  
Rather, when solving computational problems, students are provided with a MWP that already predicts the motion of some aspect of the problem (e.g., the motion of a hovercraft up to a certain point on a cliff, as seen in Fig.~\ref{fig:code}).  
This experience is similar to receiving ``user-developed'' code from a colleague and extending it to a new situation -- a common practice in science and engineering labs.  
By engaging in discussion and negotiation, students develop an understanding of what the program is doing and how it is doing it.  
They then use that understanding to modify/write additional program statements to model the situation in question.
As a result of this instruction, students having no prior experience with computational modeling are able to write essential elements of the VPython code needed to model novel situations.

We further scaffold this process by providing students with comments -- the gray ``hash-tagged'' (\#) statements in Fig.~\ref{fig:code} that are used consistently and repeatedly in each computational modeling problem.  
This common thread throughout MWPs helps to orient the students across the different instances of computational motion problems.
Typically, the number of additional program statements students will write is less than 10; they are focused on the core aspects needed to model the system (i.e., motion prediction and visualization).

For example, in the code appearing in Fig.~\ref{fig:code}, students would write several additional lines of code to model the falling hovercraft.  
This includes writing a second {\tt while}~loop that stops once the vertical position of the hovercraft coincides with the water, representing force calculations for the air drag and gravitational force in VPython code, and implementing the motion prediction algorithm that performs the appropriate Euler step.


\subsection{Content Scaffolds}\label{sec:contentscaffold}

\subsubsection{Conceptual Homework}

Much of the student experience in \pcubed is working through complex problems that require discussion and negotiation among group members to develop a complete solution.  
To support student success when solving these problems in class, we ``prime the pump'' with pre-class readings/video lectures and homework.  
Each week, students read online lecture notes, watch short video lectures, and solve conceptual and reasoning-focused homework on-line, which are meant to scaffold students' conceptual understanding that they will bring to bear in class that week.

\begin{figure}[t]
\centering
\includegraphics[width=1\linewidth]{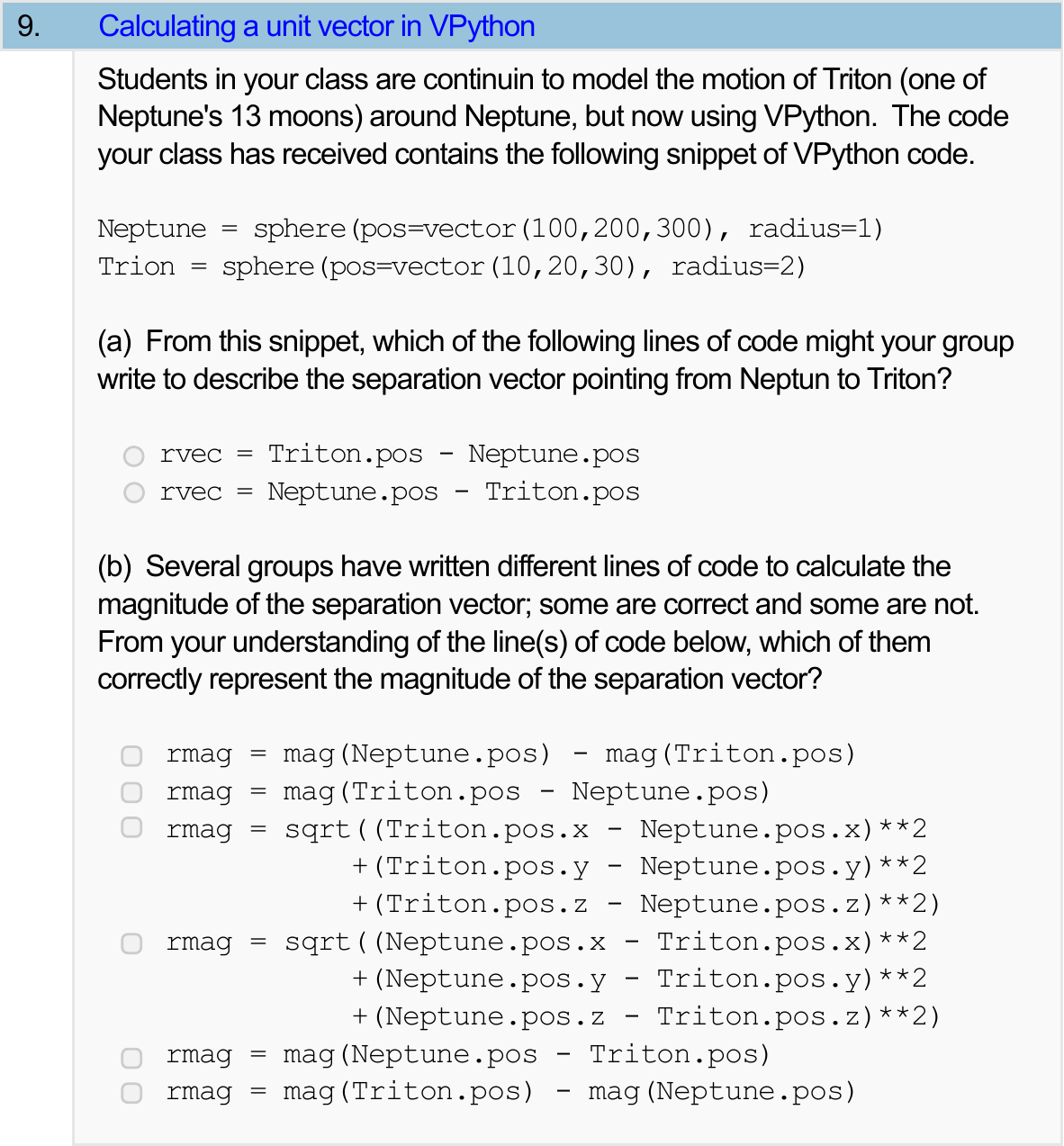}
\caption{Computational pre-homework problem focusing on the different ways to construct a unit vector in VPython.}\label{fig:compHW}
\end{figure}

For example, for in-class computational problems, the solutions can take different syntactical forms while the underlying computational algorithm is identical.  
That is, VPython has a number of functions that simplify the typing/reasoning that any group of students must undertake.  
For example, as shown in Fig.~\ref{fig:compHW}, when taking the magnitude of a vector \texttt{object.pos}, two possible methods accomplish the same goal:  the ``manual'' way of explicitly squaring, summing, and square rooting the vector components \texttt{sqrt(object.pos.x**2+obj.pos.y**2+obj.pos.z**2)} and the VPython short-cut way of \texttt{mag(obj.pos)}.  
Given the diverse computational background of students in \pcubed, we highlight these differences/similarities in the pre-homework to scaffold the discussion and solution process.

After each pair of weekly class meetings, students complete additional conceptual and reasoning-focused homework as well as more typical back-of-the-book style problems.
These are designed to provide a wrap-up of the week's material and to encourage the students to take full advantage of the opportunity to engage in group discussion and sense making in class.

\subsubsection{Tutor Questions}

With such open ended projects as shown in Sec.~\ref{sec:example}, there are many possible solution paths that lead to similarly correct solutions, and the tutors work to encourage any solution path that utilizes the concepts and ideas being focused on for that week.  
In order to ensure a group's understanding of the content and nudge a group towards productive solution pathways, we have introduced a compilation of questions that help students confront common misconceptions and test for basic understanding. As an example, for the project show in Fig.~\ref{fig:p2pa}, students are asked questions like, ``{\it What assumptions did you make about the motion of the hovercrafts?}'', ``{\it How reasonable are the values for the final speeds of the hovercrafts?}'', and ``{\it Can you group draw position and velocity vs time graphs for each hovercraft? Where do the hovercrafts meet in each graph?}''


The generation of these tutor questions is a continuous and iterative process.  
Initially, questions are generated by consulting with the literature and by ``testing'' the project with student volunteers.  
As the projects are delivered to students in-class, new solution paths inevitably arise with unique lines of reasoning.  
From these new lines of reasoning, we generate new tutor questions, which are added to the compilation.  
We aim to compile a robust set of tutor questions spanning many different solution paths.

While these questions provide tutors with general ideas about how to facilitate student work in their groups, a number of the pre-class homework mentioned previously requires an answer of the free-response type.  
Given that these homeworks are delivered using an electronic web-based system, answers may be collected easily and analyzed by tutors to highlight anything particularly difficult for that week.  
These areas of difficulty may then be used by the tutors as ``just-in-time'' style talking points during class.  
In this way, the tutor is not only informed of which topics to check for understanding on, but also on which topics students may need a little extra attention.

\section{Assessment}\label{sec:ass}

The final component of the \pcubed course is the design of the assessment. 
\pcubed students are assessed through traditional means (i.e., through homework and exams including some problems that address computational modeling concepts) as well as through some non-traditional approaches (e.g., in-class assessment) that are aligned with a focus on the development of practices (e.g., learning to problem solve and communicate effectively in a group). 



\subsubsection{Traditional Assessments}

{\bf Pre-class homework:} 
The pre-class homework scaffolds student engagement with the physics concepts for a given week as well as acting as an assessment method. 
By using conceptual questions that can be answered by reviewing the notes or pre-class videos, pre-class homework is meant to help develop students' resources \cite{Hammer2004} towards engaging with specific physics material.  
This approach focuses student attention on particular ideas (e.g., the relationship between force and velocity) that will appear in the projects that week and enables them to engage in some critical discourse around those ideas in the classroom. 
As indicated in Sec.~\ref{sec:scaf}, these questions can also take the form of questions designed to scaffold engagement with the computational projects by asking students to interpret and use lines of code that will be similar to lines of code they will see in the project that week. 

{\bf Post-class homework:} 
The purpose of the post-class homework is to provide students with a wrap up of that week's material and assess how well the students can apply the concepts they have learned about in a particular week. 
Typically, this homework focuses on one or two specific topics that were addressed by the complex problems solved in class.
However, the course is designed to build off of each week's material, so there are often call-backs to concepts dealt with in previous weeks' projects. 
The homework questions consist of multiple choice, numerical response, and graph-oriented questions. 
Both the pre-class and post-class homework are delivered, answered, and graded through an online homework system. 

{\bf Individual Portion of the Continuous Assessment and Final Exam:} 
Exams in \pcubed have both individualized and group-based sections with each section counting towards a student's overall exam grade.
The individual portion of the evening exams and the final exam are traditional hour-long exam that emphasize on both conceptual understanding (typically descriptive response questions) and numerical response type questions. 
These questions tend to be fairly standard and derivative of the post-class homework and conceptual questions that students receive in class. 
The individual portion of the exam can also include descriptive and numerical response computational questions that are computationally-focused (i.e., interpret a piece of python code and make predictions). 
Because of the emphasis on interpretation, students are given a sample code on paper, which is very similar to the type of code that they have worked with in class. 
An example of a computational exam problem is included in the appendix. 
Due to the time constraints of the exam, computational questions never require students to make physical changes to a program. 
It would be feasible to use one of the group exam problems (see Sec.~\ref{sec:trad}) as an examination of students' ability to interpret and write code, but it was decided that this would be place too much of an emphasis on computation in this course.
This individual portion of each exam makes up 75\% of a student's exam grade. 

\subsubsection{Non-Traditional, Practice-Based Assessments}\label{sec:trad}

Following the principle of constructive alignment, our assessment is aligned with our learning outcomes. 
Several of our learning goals emphasize the appropriation of scientific practices.
Hence, the design team incorporated two practice-based assessments: group exams \cite{Wieman2014} and formative feedback \cite{Shute2008,Irving2015b} based on in-class assessment.

{\bf Group exams:} 
If all of the students' contact time is working in groups where they co-construct solutions to complex problems and develop a shared understanding of the concepts, it is imperative that part of their assessment focuses on their ability to engage in such practices.\cite{Wieman2014} 
This emphasis communicates to the students that course staff expect that they will improve in these areas and that what students are doing in class will be assessed. 
The group exams are the second part of the continuous assessment and final exam.
They are open-ended, hand graded exams completed by the groups using the same resources available to them during class (i.e. they can use a laptop or notes). 
Group exams follow the same format as the problems experienced in class except that the exam problems are simpler due to the time constraint (i.e., 1 hour for group exams). 
The students produce a solution in the form of a report that outlines the model world that they have constructed as well as the approach that they have taken to grapple with the presented scenario. 
A group's report is graded using a rubric that emphasizes how well the students constructed the model world, the predictions they made within the model world, and how they would make changes to their model world (based on these predictions) to make it more physically realistic. 
In addition to a numerical score, students are given feedback on their report.
This feedback discusses which aspects of the rubric that they did well on, which aspects they did not pay sufficient attention to, and how they might devote more attention to areas their solution lacked in the future. 

The provision of feedback is to attempt to establish new expectations after each group exam. 
This negotiation of new expectations is an acknowledgment that this type of assessment is a new experience for the majority of students taking the class.
Therefore, our initial expectations have to be set at an achievable level and then adjusted to allow for growth. 
For example, the expectations for how students reflect on their model world are initially small but then adjusted upward as the feedback communicates this new expectation and how a group might achieve a better score. 
The low initial expectations stem from our understanding that this type of metacognitive reflection needs to be developed over time as well as that the reflection comes at the end of the solution process. 
This means that in the group exam, groups (especially in their first group exam) may have not developed adequate, effective practices in time management.
Thus, groups might not afford themselves enough time to engage in such reflection in any meaningful way at the end of the exam. 
A student's score on a group exam comprises 25\% of the total score on each exam.

{\bf Formative feedback based on in-class assessment:} 
The \pcubed learning environment is designed so that the majority of the physics learning occurs while the students work in small groups within the classroom. 
\pcubed students rely on each other to develop their understanding of course material including computational elements. 
Under the principle of constructive alignment,\cite{Biggs2011} a learning activity should be designed so that there are clear, related learning outcomes for the activity and appropriate assessment for giving feedback to the learner. 
To provide this feedback and to motivate students to engage with the in-class project activities, the design team constructed the an assessment that would be based on in-class work. 
Each week, we assess students on their group work and provide them with feedback for improvement. 
In particular, we provide students with individualized feedback based on the following three criteria:

\begin{itemize}[noitemsep,nolistsep]
  \item How well do you develop your own understanding of the physics (Individual Understanding)? (e.g., how well prepared was the student for class?)
  \item How well does your group ensure all members develop an understanding of the physics (Group Understanding)? (e.g.,  is the student openly expressing their ideas when they are confused for group discussion? is the student tutoring other students when incorrect/incomplete ideas are presented?)
  \item How well does your group manage itself in terms of the discussion and use of ideas (Group Focus)? (e.g., are everyone's ideas being respected and discussed?)
  \end{itemize}

Tutors provide the students with written feedback before the start of each new project. 
Fig.~\ref{fig:aoifefeedback} is an example of the type of feedback the students would receive during a semester.

\begin{figure}[t]
\fbox{
\begin{minipage}{0.95\columnwidth}
\flushleft{
{\bf Week 7 Feedback: Aoife}\\\vspace*{10pt}

Again, your group did a great job working together. You all worked through both problems well and had good discussions about the physics in the problems where it was clear that each person's ideas were respected and valued. This week it's clear that you all listened to the feedback as Sean, Niamh, and Cliona were all contributing to the intellectual work in the problem most of the time. You seemed to be included for most of the work, but sometimes it seemed like they were working on something and then catching you up. Maybe shake up the seating order to put yourself in the middle so you can contribute more; your questions help your group think about problems in new ways. So next week we'd like to see you try to insert yourself into the group work throughout the week. Also, Aoife, we'd really like to see you question different aspects of the work -- you seem to have a critical eye when something is fishy, but you don't often express it. But, y'all did a great job again!\\\vspace*{10pt}

{\bf Group Understanding (out of 100):} $87.5$\\
{\bf Group Focus (out of 100):} $93.75$\\
{\bf Individual Understanding (out of 100):} $75$\\
{\bf Weekly Group Work Score (out of 4):} $88.5$
}
\end{minipage}
}

\caption{Sample feedback received by students in P$^3$.}
\label{fig:aoifefeedback}
\end{figure}

Instructor feedback is based on the previous weeks' project performance and focuses on one type of participation that they excelled at and one type to work on in the next project. 
The feedback provides students with suggestions on how they might go about achieving that improvement. 
They also discuss group functioning as a whole to attempt to facilitate good group dynamics.
This feedback has been an essential part of our design. 
It has communicated that we value the development of group orientated practices (e.g. debating personal understanding) and has also acted as another scaffold that provides students with the support to begin to reflect on how they are performing in their group. 
From the perspective of the computational modeling problems, this feedback has been critical to our success -- ensuring that less computationally-prepared students take charge of writing the programs as well as promoting more computationally-prepared students to tutor their classmates on different aspects of computational modeling. 
This group structure ensures that students discuss and negotiate computational modeling concepts while writing programs to model authentic physical situations. 
As with the group exams, the formative feedback is multi-modal; it provides a scaffold for the development of multiple practices, explicitly communicates our expectations in regards the development of practices, and assesses such development. 

\section{Learning Outcomes}\label{sec:learningout}

{\subsection{Conceptual Learning}}

To evaluate force and motion learning outcomes, we used a traditional pre-post assessment method that is typical of most curricular and pedagogical evaluations.\cite{Hake1998,Caballero:2012ee}
At MSU, it is traditional to take this type of data for our introductory, calculus-based mechanics courses via the Force Motion Conceptual Evaluation (FMCE). \cite{Thornton1998}
The results of this pre/post testing are often presented in the form of a normalized gain, which is the average increase in students' scores divided by the average increase that would have resulted if all students had perfect scores on the post-instruction test. 
In traditional lecture-based instruction at MSU, normalized gains on the FMCE range from 0.10-0.35.
For the most recent offerings of \pcubed (N = 160), students (on average) earn normalized gains of 0.60, which is comparable to other transformed learning environments.
For example, the University of Colorado Boulder (CU) have previously presented a range of normalized gains for their various interactive engagement introductory mechanics classes from 0.32 to 0.64.\cite{Finkelstein2005a} 
Both the CU and MSU data aligns with Hake's widely referenced plot,\cite{Hake1998} which although completed with FCI data is still analogous to the normalized gains that should be expected for a transformed interactive engagement classroom. Therefore, the MSU P$^3$ score meets expectations for an interactive engagement classroom.


{\subsection{Attitudes and Beliefs}}

One of the fundamental building blocks of the P$^3$ design is to engage students in practices that are authentic to physicists (e.g., Sec.~\ref{sec:model}). 
A design goal relevant to this course design principle is to facilitate a change in the attitude of the students towards physics through engagement with these authentic science practices. 
This is not an explicit learning goal communicated to the students and so could be considered part of the \pcubed ``hidden curriculum.''\cite{redish2010introducing} 
The Colorado Learning Attitudes about Science Survey (CLASS) \cite{Adams2006} is an evaluation of students' cognitive attitudes toward the nature of physics and the practice of physics, which also indicates whether students have grown to like or dislike physics. 
Like the FMCE, the CLASS is a pre/post assessment in which changes in students' ``attitudes and beliefs" from the beginning of the semester are compared to those at the end of the semester. 
The general outcome within the physics community of these attitudinal surveys to date has been a consistent negative shift in students' attitudes after taking an introductory physics courses with a few exceptions.\cite{Finkelstein2005a,Traxler2015}
This negative shift is a consistent result among all introductory lecture courses (regardless of instructor or level).
However, students taking \pcubed display on average a .05\% positive shift. 
Although this shift is relatively low, within the context of other transformed introductory classrooms, it is a result worth noting given the negative shift normally experienced.


{\subsection{Students' reflections on \pcubed}}

Obtaining student buy-in is a crucial step to the success of any transformed classroom.\cite{dancy2008barriers}
At the end of the semester of the first two iterations of \pcubed we interviewed students about various aspects of their experiences in taking the transformed class. 
These interviews attempted to assess student buy-in, to develop improvements based on the students' thoughts about the class, and to observe self-development in the more practice and affective-focused learning goals. 
The following paragraphs highlight some of the common themes regarding student reflection on these learning goals. 

{\subsubsection {Change in perception of utility of computers for doing science}}

By including computation in the classroom (Sec.~\ref{sec:goals}), we did not intend for students to learn a programming language.
Instead, we attempted to create a learning environment in which students can interact with computational models (in a scaffolded way) to learn physics through computation as well as to encourage the students to begin to see the utility of computational modeling in their future work. 
The following extract from an interview indicates the growth of the utility perception as a result of the students' interactions in the class.

\begin{description}
\item[Student] {\it I could see benefits...I work in the analytical sciences department, which we do a lot of sample testing and work with people who do failure analysis, and failure analysis would be great to do with something like this (code) to see what happens over time. I'm not sure how you would program that in, but I'm sure that you can. I think that would be very beneficial, in a presentation, of what changing a product, what could happen to that product over time.}
\item[Interviewer] {\it And is this something you would have thought of before you had this experience?}
\item[Student] {\it Nope, I would have just thought, let's experiment on it and we can show a trend line...for me if someone brought a presentation and they ran this code and it showed me a visual representation of what happened to that plate, I would take that more to heart than just looking at a trend line on a graph.}
\end{description}

From this extract, we can observe this student is beginning to see computation as a tool that can be applied in multiple contexts and not just to make predictions about hovercrafts falling off cliffs. 
We are not saying that all of the students interviewed exhibited such changes in perception. 
Several of the students interviewed did not understand why computation was being used in the classroom nor what advantages it was affording them. 
In the future, this will be addressed by making sure to not only integrate computation into projects but by also communicating ``why" it is part of the design of the classroom.

{\subsubsection{Affect and Motivational Influence of \pcubed}}

The \pcubed learning environment is designed to encourage students to think of themselves as part of a community of learners who all have the same intended goal of learning physics. 
The fact the students considered the classroom as a community and the effects of being a member of this community was evidenced in nearly all of the student interviews conducted. 
The overall sentiment was that the students enjoyed collaborating in their groups:

\begin{description}
\item[Student] {\it I kind of like coming to class just because I like my groups, and I like talking to them, and I like working with them, so I think my idea about physics has changed. I see it more in a positive light.}
\end{description}

And that the students began to see each other as resources for learning in a community that will help each other:

\begin{description} 
\item[Student] {\it With \pcubed we all had each other's phone numbers, we could talk to each other, or Facebook or email whatever, and so I could email the other person and be like I don't understand this can you explain?}
\end{description}

Not only was it a community in which students viewed each other as resources, but it was also a community where no one was judged for not understanding a concept or idea, and instead questions are encouraged:

\begin{description} 
\item[Student] {\it I can ask any question and not feel like an idiot...I swear I asked like five times, and everyone explained to me over and over again what it was until I got it, like, I didn't feel judged at all, it was really nice.}
\end{description}

But there are certain responsibilities and expectations that come with membership within a community, which has a common goal and engages in shared practices. 
For example, the student in the next extract talks about the stress associated with being prepared for \pcubed class sessions:

\begin{description}
\item[Student] {\it The stress of going into \pcubed, you have to be prepared going in, and if your not, it's almost as if that stress made me want to learn more because I wanted to be prepared and I wanted to be like a badass in my group.}
\end{description}

Of course, there is a certain amount of stress associated with any class that is taken, but this element of stress would be considered as a negative affective factor when it comes to classroom design. 
But not all negative affective factors are bad for a student just as not all positive affective factors are useful for a student. 
In this particular case, the stress comes from a student not wanting to let their group down, which results in the positive practice of more preparation so that they can contribute to their group and learning community. 
Overall, the narrative that these extracts indicate is that \pcubed develops as a learning community, which was an important design goal about students' feeling that they are engaging in a community of practice.

{\subsubsection {Improved Group Skills}}

Finally, we designed \pcubed to improve the ability of a student to collaborate and problem solve effectively in a group. 
All of the class time is focused on group learning (Sec.~\ref{sec:example}) and one of the purposes of the formative feedback (Sec.~\ref{sec:trad}) is to help students improve their capacity to work in a group. 
Two themes emerged from the data in regards to the group aspect of the class design: (1) students saw the class as trying to improve their ability to be effective in a group and (2) this emphasis on developing group skills makes a lot of sense because the students perceive that they will have to work effectively in groups in the future. 
The first point is evidenced in the following quote:

\begin{description}
\item[Student] {\it I think because a main point of the class was not just exposing you to the physics and getting you good at the physics but also working in groups and being good on a team.}
\end{description}

The majority of students who have taken \pcubed to date have been various engineering majors. 
Engineering like science is a discipline that is built around projects that are worked on as a team. 
The engineering students in \pcubed particularly appreciated the emphasis that was being placed on group work as they were aware of the substantial role it will play in their future careers:

\begin{description}
\item[Student] {\it This physics class is mainly for scientists and engineers. And engineer[ing] is one hundred percent working with someone else, you're not going to be an engineer and do something on your own, your going to have to show it to somebody or explain it to somebody, so in this it's good because we have to explain what we are thinking to other people.}
\end{description}

To date, we have not developed a research instrument to measure the development of a student's ability to work in a group, but students to seem to be receiving a message from \pcubed that group work is important. 
Anecdotal evidence based on the observations of multiple tutors as well as the group exam reports, which students turn-in, seem to indicate a positive progression in our students' ability to work in groups.

{\section{Sustainability and adaptability of the \pcubed transform}}

Integrating the opportunity for students to engage in authentic practice is important for future scientists and engineers to appropriate a grounding in practices that are fundamental to ``doing" science and engineering in a professional capacity (e.g., ``pathway practices''). 
It can be argued that exposure to scientific practices occurs in undergraduate laboratory experiences, but this hides the fact that the appropriation of practices is not explicitly supported and instead assumed due to the context. 
Integrating computation into the introductory physics curriculum is a foundational first bridge that allows access to multiple scientific practices.
Computational modeling is an especially important practice for future scientists and engineers to learn as conducting modern science and engineering requires using computational tools. 
 In this paper, we have demonstrated one model for integrating authentic practice into the introductory physics curriculum that relies on the alignment between learning goals (Sec.~\ref{sec:goals}), opportunity to engage in practices (Sec.~\ref{sec:example}), learning scaffolds (Sec.~\ref{sec:scaf}), and assessment (Sec.~\ref{sec:ass}).

This model has been designed to allow for some affordances for potential adopters and their students. 
For example, \pcubed does not require faculty to provide any formal computational instruction due to its group-based nature nor does it overwhelm the students by asking too much in regards to the appropriation of computational skills. 
By focusing on providing the opportunity for students to experience the utility of computation for themselves and learning some basic computational skills instead of developing fluency with a computational language, we circumvent overloading the students with unobtainable expectations and learning goals. 
We have also taken logistical steps to alleviate concerns about students' potential struggles with computational modeling by placing a student with some computational background in each of the four-person groups at the beginning of the semester. 
We also believe that another affordance is the variation in the degree of adoption that is available for potential adopters. 
By detailing the different scaffolds (Sec.~\ref{sec:scaf}) and assessments (Sec.~\ref{sec:ass}) and how they relate to each other, we aim to help adopters pick and choose elements of the model that would be a good fit for their particular context. 

However, this model is not without its shortcomings. 
To focus on all the practices mentioned above while also assessing if students are engaging with them in a meaningful way, there is a need for someone to assess what the students are doing. 
The initial implementation of \pcubed involved two faculty members, one post-doc, and a graduate student. 
This allocation of staff is an unrealistic model for sustainability at almost any institution. 
To make this model of instruction sustainable past the initial funding, we have implemented a learning assistant (LA) program.\cite{Otero2010} 
The personnel model that is sustainable for MSU going forward is a single faculty member, a graduate student, and ten undergraduate LAs for a one hundred student section. 
The LAs are all drawn from students who have come through the class. 
These LAs go through training at the beginning of the semester, but are already well equipped to give formative feedback and to guide groups through problems as they have already experienced the course from the student perspective.
By altering aspects of the course design and what adopters choose to assess, fewer instructional staff would be needed. 
An second shortcoming is by refocusing to student-centric activities, there is a loss of content, which is typical for such an introductory physics transformation. 
The content that was left out did not fit with the narrative of the course with the focus on the main principles (see Sec.~\ref{subsec:content}). 
We deemed this an acceptable loss due to the arguments presented in The Framework\cite{NRC2011} and the high value that both faculty and students placed on learning scientific practices.

The final component of any curriculum design is further iteration to ensure sustained quality. 
However, for this iteration to occur for the \pcubed environment, research has to catch up and inform future decisions around key elements of the design. 
For example, we need to understand how structuring computational modeling instruction around providing an experience to change students' perceptions of the utility of computers as opposed to learning a particular computational language affect how and what students are learning. 
Investigating the long-term effects of formative feedback and how it influences the appropriation of physics practices is also an open question. 
As mentioned, there is no inventory to assess the development of group skills or in essence any of the other practices that are being focused on to make any concrete claims about the effect of taking \pcubed. 
Future work by our group and the physics education research community, more broadly, should investigate these gaps in knowledge to develop a more complete understanding of how to integrate content and practice successfully.

\begin{acknowledgments}
This work has been financially supported by the CREATE for STEM Institute (funding for curriculum development and staffing) and the Department of Physics and Astronomy (learning assistant support) at Michigan State University. The authors would like to thank the faculty who have worked in the courses: Tyce DeYoung, Richard Hallstein, Lisa Lapidus, and Stuart Tessmer.
\end{acknowledgments}


\bibliography{AJPcubed}
\bibliographystyle{apsper}

\end{document}